# Formal Analysis of an Authentication Protocol Against External Cloud-Based Denial-of-Service (DoS) Attack

Marwan Darwish, Abdelkader Ouda, Luiz Fernando Capretz

*Department of Electrical and Computer Engineering*
*University of Western Ontario*
*London, Canada*
*{mdarwis3, aouda, lcapretz}@uwo.ca*

## Abstract

The Denial-of-service (DoS) attack is considered one of the largest threats to the availability of cloud-computing services. Due to the unique architecture of cloud-computing systems, the methods for detecting and preventing DoS attacks are quite different from those used in traditional network systems. A main target for DoS attackers is the authentication protocol because it is considered a gateway to accessing a cloud's resources. In this work, we propose a cloud-based authentication protocol—one that securely authenticates the cloud's user and effectively prevents DoS attack on the cloud-computing system—by involving the user in a high computation process. Then, we analyze the protocol via Syverson and Van Oorschot (SVO) logic to verify the authentication process of the protocol in a cloud-computing system.

## 1. Introduction

Cloud computing is the utilization of a combination of hardware and software to provide services to end users over a network (e.g. the Internet). It includes a set of virtual machines that simulate physical computers and provide services, such as operating systems and applications. However, configuring virtualization in a cloud-computing environment is critical. A cloud-computing structure relies on three service layers: Infrastructure as a Service (IaaS), Platform as a Service (PaaS), and Software as a Service (SaaS) (Fig. 1). IaaS gives users access to physical resources, networks, bandwidth, and storage. PaaS builds on IaaS and gives end users access to the operating systems and platforms necessary to build and develop applications, such as databases. SaaS provides end users with access to software applications.

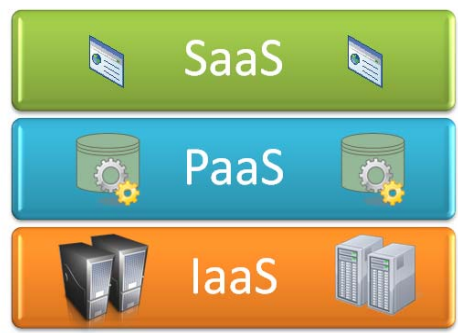

Figure 1. Cloud-Computing Layers

DoS attacks are major security risks in a cloud-computing environment because resources are often shared by many users. A DoS attack targets resources or services in an attempt to render them unavailable by flooding system resources with heavy amounts of artificial traffic. The objective of DoS attacks is to consume resources—memory, CPU processing space, or network bandwidth—in order to make them inaccessible to end users by blocking network communication or denying access to services. Handling DoS attacks at all layers in cloud systems is a major challenge due to the difficulty of distinguishing an attacker's requests from legitimate user requests.

Detecting a DoS attack in its early stage, in the upper layer (SaaS), is an ideal approach to avoid the destruction caused by DoS attacks on other layers. However, all service requests for SaaS need to be authenticated in order to operate. Verifying users via an authentication protocol is an initial stage in accessing these systems. Consequently, the authentication protocol is a main target of attackers implementing a DoS attack, and decreases the availability of cloud services. The use of existing strong authentication protocols from traditional network systems in cloud-based applications may lead to DoS attack vulnerability. This is because the initiation of a massive amount of authentication processes could exhaust the cloud's resources, and make the cloud-based application unreachable.

In this paper, we discuss the types of possible external DoS attacks in a cloud-computing environment. Then, we propose an authentication protocol against DoS attack, and present a formal analysis of the proposed protocol. Section 2 provides an overview of DoS attacks. Section 3 states the need for authentication protocols. Section 4 describes a proposed cloud-based authentication protocol against external DoS attacks. Section 5 analyzes the proposed protocol via Syverson and Van Oorschot (SVO) logic. Finally, Section 6 presents a brief summary of the paper.

## 2. DoS Overview

DoS attacks have become more sophisticated in recent years. Many websites and large companies are targeted by these types of attacks. The first DoS attack was reported in 1999 [1]. In 2000, large resource companies, including Yahoo, Amazon, CNN.com and eBay, were targeted by DoS attacks and their services were stopped for hours [2]. Register.com was targeted by a DoS in 2001; this was the first DoS attack to use DNS servers as reflectors [3]. In 2002, service disruption was reported at 9 of 13 DNS root servers due to DNS backbone DoS attacks. This attack recurred in 2007 and disrupted two DNS root servers. In 2003, Microsoft was targeted by a DoS called Worm Blaster. One million computers were attacked by MyDoom in 2004. In 2007, a DoS attack was carried out by thousands of computers, and targeted more than 10,000 online game servers. In 2008, a DoS attack targeted Wordpress.com and caused 15 minutes of denial [4]. In 2009, a cloud-computing provider named GoGrid was targeted by a large DoS attack, and approximately half of its thousands of customers were affected. In 2009, Register.com was affected again by a DoS attack. In the same year, some social networking sites, including Facebook and Twitter, were targeted by a DoS. Many websites were attacked by DoS in 2010, including the Australian Parliament House website, Optus, Web24, Vocus, and Burma's main Internet provider. In 2011, Visa, MasterCard, PayPal, and PostFinance were targeted by a DoS that aimed to support the WikiLeaks founder [4]. In the same year, the site of the National Election Commission of South Korea was targeted by DoS attacks. Furthermore, thousands of infected computers participated in a DoS attack that targeted the Asian E-Commerce Company in 2011 [4]. In 2012, the official website of the Office of the Vice President of Russia was unavailable for 15 hours due to a DoS attack [4]. In the same year, many South Korean and United States (US) websites were targeted by DoS. Godaddy.com websites reported service outages because of such an attack. In 2012, major US banks and financial institutions became the target of a DoS attack. DoS attacks are evolving rapidly and are targeting large companies, which cause huge financial losses to those companies and websites globally.

DoS attacks affect all layers of the cloud system (IaaS, PaaS, and SaaS) and can occur internally or externally. An external cloud-based DoS attack starts from outside the cloud environment and targets cloud-based services. This type of attack affects the availability of services. The most affected layers in the cloud system by an external DoS attack are the SaaS and PaaS layers.

The two categories of cloud-based DoS attacks are internal and external cloud-based DoS [5]. Descriptions of external cloud-based DoS attacks are presented in the following sections.

### 2.1. IP spoofing attack

In the Internet Protocol (IP) spoofing attack, packet transmissions between the end user and the cloud server are intercepted and their headers modified such that the IP source field in the IP packet is forged by either a legitimate IP address, as shown in Fig. 2, or by an unreachable IP address. As a result, the server will either respond to the legitimate user machine, which affects the legitimate user machine, or the server will be unable to complete the transaction to the unreachable IP address, which affects the server resources. Tracing such an attack is difficult due to the fake IP address of the IP source field in the IP packet.

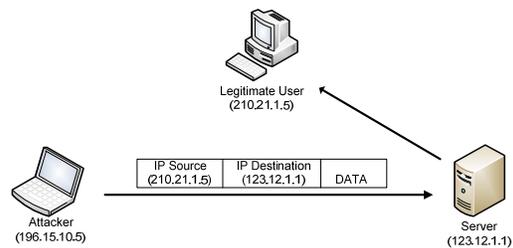

Figure 2. IP spoofing attack

## 2.2. SYN flooding attack

A Transmission Control Protocol (TCP) connection starts with a three-way handshake, as shown in Fig. 3(a). A typical three-way handshake between a legitimate user and the server begins by a connection request sent from the legitimate user to the server in the form of a synchronization (SYN) message. Then, the server acknowledges the SYN by sending back (SYN-ACK) a request to the legitimate user. Finally, the legitimate user sends an ACK request to the server to establish the connection. SYN flooding occurs when the attacker sends a huge number of packets to the server but does not complete the process of the three-way handshake. As a result, the server waits to complete the process for all of those packets, which makes the server unable to process legitimate requests, as shown in Fig. 3(b). Also, SYN flooding can be accomplished by sending packets with a spoofed IP address. A sniffing attack is also considered a type of SYN flooding attack. In a sniffing attack, the attacker sends a packet with the predicted sequence number of an active TCP connection with a spoofed IP address. Thus, the server is unable to reply to that request, and the resource performance of the cloud system is then affected.

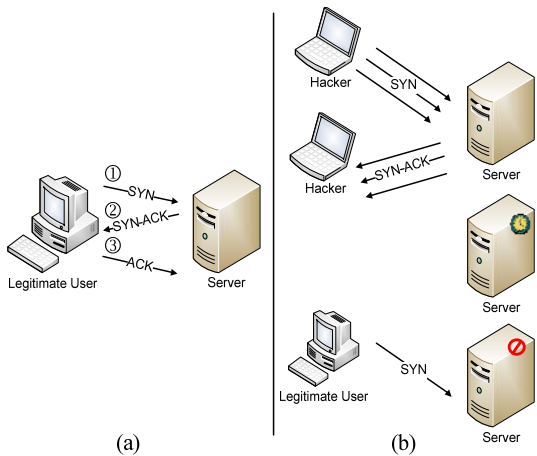

Figure 3. SYN flooding attack

## 2.3. Smurf attack

In a smurf attack, the attacker sends a large number of Internet Control Message Protocol (ICMP) echo requests. These requests are spoofed such that their source IP address is the victim's IP, and the IP destination address is the broadcast IP, as shown in Fig. 4. As a result, the victim will be flooded with broadcasted addresses. The worst case occurs when the number of hosts who reply to the ICMP echo requests is too large.

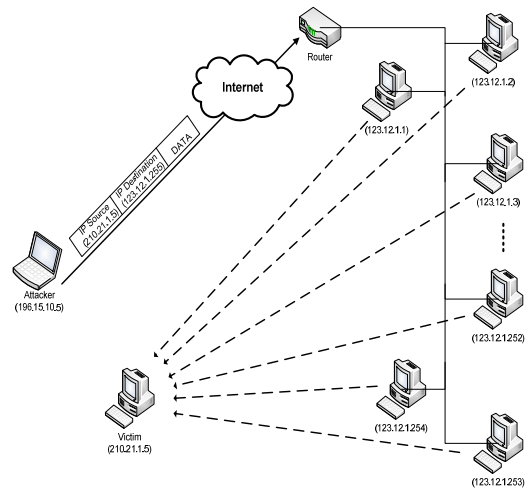

Figure 4. Smurf attack

## 2.4. Ping of death attack

In the ping of death attack, the attacker sends an IP packet with a size larger than the limit of the IP protocol (>65,535 bytes). Handling an oversized packet affects the victim's machine within the cloud system as well as the resources of the cloud system.

## 3. The Need for Authentication Protocols

There are a number of authentication protocols that are strong enough to verify identities and protect traditional networked applications. However, these authentication protocols may introduce DoS risks when adopted in cloud-based applications. This is due to the utilization of a heavy verification process that may consume the cloud's resources and disable the application service. The OAuth protocol [6] is currently a widely used authentication protocol that controls the access of third-party applications to an HTTP service. In OAuth, the resource owner can allow a third-party client to access its resources through the owner. However, any insecure implementation of OAuth protocol can lead to the possibility of a DoS attack.

Many authentication protocols have been proposed for the SaaS layer, but they do not protect against DoS attacks Yassin et al. [7] proposed an authentication process that uses a one-time password (OTP) with mutual authentication of the user and cloud server. Yassin's authentication schemes defended against replay attacks, but not DoS attacks. Some cloud-based authentication protocols for DoS prevention have been proposed (e.g., Choudhury et al. [8], Hwang et al. [9], Jaidhar [10], and Tsaur et al. [11]), but they use a smart card reader for the authentication process. Furthermore, the Yassin et al. [12] scheme also recommend the use of an extra physical device such as a fingerprint scanner.

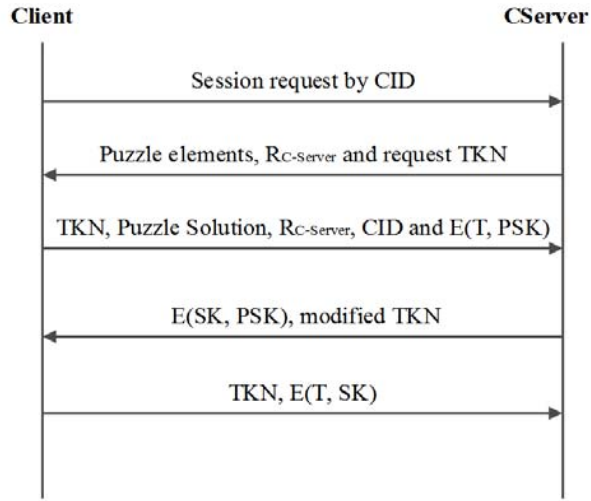

Figure 5. A proposed authentication protocol

On their own, the authentication protocols may lead to DoS attack vulnerability. Therefore, it is necessary and important to verify DoS resistance in every process of the authentication protocol. For instance, verifying a huge number of signed messages via the server consumes the resources of the server to a significant degree, particularly when the attacker sends a massive number of forged signed messages. On the other hand, sending a typical client credential with each request in the authentication protocol will force the server to verify these requests based on the stored information at the server. As a consequence, server resources will be exhausted when dealing with a large number of requests.

In this work, we propose a cloud-based authentication protocol that is capable of defending against external DoS attackers. The proposed protocol protects from DoS attacks by increasing the efforts made by the requestor. This protocol is capable of distinguishing between a legitimate user's requests and a DoS attacker's requests. The proposed authentication protocol was designed to be completely aware of any risk of DoS attacks, and is described in the following section.

## 4. A Proposed Authentication Protocol Against External DoS Attack

Table (1) shows the notations that are used in the proposed authentication protocol.

Table 1. Notations of the authentication protocol

| Notation | Description |
| --- | --- |
| Client | The cloud client |
| CServer | The cloud server/service provider |
| CID | Cloud user (Client) ID |
| TKN | Token; the key of the token is known only by CServer |
| SK | Session key |
| $R_{CServer}$ | The nonce that is generated by CServer |
| T | Time stamp |
| PSK | Pre-shared key |
| MK | Master key of CServer |

The proposed authentication protocol, as shown in Fig. 5, starts when the client sends a request for a service with CID to CServer. CServer will reply directly to the client by sending puzzle elements to keep the client's resources busier and increase the participation cost of the client's side. At this point, CServer will block any CID that has performed three consecutive requests within a low time threshold to prevent DoS attacks. However, if the attacker attempts to launch a DoS attack by sending requests with randomly generated CID, the resources of CServer will be less affected because it will simply reply to each request with puzzle elements, and it will not check for the client's data on any related database system or on any stored information. As a result, CServer is not required to check each request's information to make any decision about the request. Thereafter, CServer will ask the client for the puzzle solution. In addition, CServer will send a nonce $R_{CServer}$, and request the token (TKN) from the client. The client should have the TKN from the time when the client registered to the cloud system. Moreover, the key of the TKN is known only by CServer. Once the client solves the puzzle, the client will send the TKN, puzzle solution, the received $R_{CServer}$, CID, and the encrypted timestamp T by a pre-shared key (PSK) to CServer for validation. At this point, CServer will have all the information required to validate the authentication requests, so CServer can apply the validation process to only a few operations, such as the following:

- CServer will check the solution of the puzzle problem to determine whether or not it is correct.

- CServer will check the time difference between the received timestamp T and the current timestamp to determine whether or not it is a reasonable time difference in which to find the solution.

If either of the two conditions does not apply, CServer will drop the request and consider it as an attacker's request. If the client request passes the two conditions, CServer will decrypt the TKN and validate the decrypted information that contains the CID. After the validation process, CServer will generate the session key (SK), which is encrypted via a pre-shared key (PSK). Moreover, CServer will add both the SK and T information to the TKN. Consequently, CServer is not aware of possible DoS attacks on key storage. Furthermore, CServer can apply the refreshment property of the session key for future communication by adding the SK information to TKN. Then, CServer will send the generated SK that is encrypted by PSK, and the modified TKN to the client. Finally, the client will confirm the SK by sending back the received TKN, and the encrypted T by the SK. Then, CServer will decrypt the TKN, validate the CID and T, and then confirm the SK. Later, the two parties can agree upon the sub-session keys by re-applying the last two processes so that the CServer can generate a sub-session key and add it to the TKN without storing it in the cloud system.

## 5. Formal Analysis of the Proposed Authentication Protocol via SVO Logic

Burrows, Abadi and Needham (BAN) proposed a "belief logic" [13] to analyze the security requirements of the authentication protocols. Subsequently, SVO logic was introduced by Syverson et al. [14] as an extension model to cover some limitations of BAN. SVO uses some of its own notations in addition to those used in BAN. The analysis steps of any authentication protocol via SVO [15] are as follows:

1) **Initial assumption** is an assumption of initial status of the protocol such as belief on freshness of nonces, and so on.
2) **Received message assumption** is an assumption about messages each party receives in the case that a protocol completes faithfully.
3) **Comprehension assumption** is an assumption about what is comprehended by the principal of each received message.
4) **Interpretation assumption** is an assumption about the way each party interprets the received messages.
5) **Derivation** derives the beliefs that each party obtains by the previous assumptions, and checks the authentication goals that are derived.

Table (2) shows the notations that are used in SVO logic.

Table 2. Notations of the SVO

| Notation | Description |
|---|---|
| $P$ believes $X$ | $P$ can take $X$ as true |
| $P$ received $X$ | $P$ has received a message containing $X$ |
| $P$ said $X$ | $P$ believes $X$ when $P$ sent it |
| $P$ says $X$ | $P$ has said $X$ |
| $P$ has $X$ | $X$ is initially available to $P$, freshly generated by $P$, or received by $P$ |
| $P$ controls $X$ | $P$ has a jurisdiction on $X$ |
| $fresh(X)$ | $X$ is fresh, and it has not been sent before |
| $P \xleftrightarrow{k} Q$ | $P$ and $Q$ communicate with each other by a good shared key $k$ |
| $PK_\psi(P, k)$ | $k$ is a public encryption key of $P$. Only $P$ can read messages encrypted by $k$ |
| $PK_\sigma(P, k)$ | $k$ is a public signature key of $P$. The key $k$ verifies that the messages signed by the corresponding private key $k^{-1}$ are from $P$ |
| $PK_\delta(P, k)$ | $k$ is a public key-agreement key of $P$. A Diffie-Hellman key formed with $k$ is shared with $P$ |
| $\{X\}_k$ | $X$ encrypted under key $k$ |
| $\lfloor X \rfloor_k$ | $X$ signed with key $k$ |
| $<X>_{*P}$ | The received message $X$ is unrecognized by $P$ |

The two inference rules of SVO are:

Modus Ponens

$$\frac{\varphi \quad \varphi \rightarrow \psi}{\psi}$$

Necessitation

$$\frac{\vdash \varphi}{\vdash P \text{ believes } \varphi}$$

Where $\Gamma \vdash \varphi$ means that $\varphi$ can be derived from the set of formulas $\Gamma$. Using the above rules "$\vdash \varphi$" means that $\varphi$ is a theorem.

The axioms of SVO are:

- Belief Axioms

A1. $(P \text{ believes } \varphi \wedge P \text{ believes } (\varphi \rightarrow \psi)) \rightarrow P \text{ believes } \psi$

A2. $P \text{ believes } \varphi \rightarrow \varphi$

A3. $P \text{ believes } \varphi \rightarrow P \text{ believes } (P \text{ believes } \varphi)$

A4. $\neg(P \text{ believes } \varphi) \rightarrow P \text{ believes } (\neg P \text{ believes } \varphi)$

- Source Association Axioms

A5. $(P \stackrel{k}{\leftrightarrow} Q \wedge R \text{ received } \{X \text{ from } Q\}_k) \rightarrow (Q \text{ said } X \wedge Q \text{ has } X)$

A6. $(PK_\sigma(Q,k) \wedge R \text{ received } X \wedge SV(X, k, Y)) \rightarrow Q \text{ said } Y$

- Key Agreement Axioms

A7. $(PK_\delta(P, k_P) \wedge PK_\delta(Q, k_Q)) \rightarrow P \stackrel{F_0(k_P,k_Q)}{\longleftrightarrow} Q$

A8. $\varphi \equiv \varphi [F_0(k, k') / F_0(k', k)]$

- Receiving Axioms

A9. $P \text{ received } (X_1, \dots X_n) \rightarrow P \text{ received } X_i$ for $i = 1, \dots, n$

A10. $(P \text{ received } \{X\}_{k+} \wedge P \text{ has } k^-) \rightarrow P \text{ received } X$

A11. $(P \text{ received } \lfloor X \rfloor_k) \rightarrow P \text{ received } X$

- Possession Axioms

A12. $P \text{ received } X \rightarrow P \text{ has } X$

A13. $P \text{ has } (X_1, \dots X_n) \rightarrow P \text{ has } X_i$ for $i = 1, \dots, n$

A14. $(P \text{ has } X_1 \wedge \dots \wedge P \text{ has } X_n) \rightarrow P \text{ has } F(X_1, \dots X_n)$

- Comprehension Axiom

A15. $P \text{ believes } (P \text{ has } F(X)) \rightarrow P \text{ believes } (P \text{ has } X)$

- Saying Axioms

A16. $P \text{ said } (X_1, \dots X_n) \rightarrow P \text{ said } X_i \wedge P \text{ has } X_i$ for $i=1,\dots, n$

A17. $P \text{ says } (X_1, \dots, X_n) \rightarrow (P \text{ said } (X_1, \dots X_n) \wedge P \text{ says } X_i)$ for $i = 1, \dots, n$

- Freshness Axioms

A18. $\text{fresh } (X_i) \rightarrow \text{fresh } (X_1, \dots, X_n)$ for $i = 1, \dots, n$

A19. $\text{fresh } (X_1, \dots, X_n) \rightarrow \text{fresh } F(X_1, \dots, X_n)$

- Jurisdiction and Nonce-Verification Axioms

A20. $(P \text{ controls } \varphi \wedge P \text{ says } \varphi) \rightarrow \varphi$

A21. $(\text{fresh } (X) \wedge P \text{ said } X) \rightarrow P \text{ says } X$

- Symmetric Goodness Axiom

A22. $P \stackrel{k}{\leftrightarrow} Q \equiv Q \stackrel{k}{\leftrightarrow} P$

In the proposed protocol, the initial three processes are computationally intensive, with exponential complexity ($O^{2n}$), to ensure that the client expends some effort. The next two processes of the protocol authenticate the client, as shown in Fig. 6.

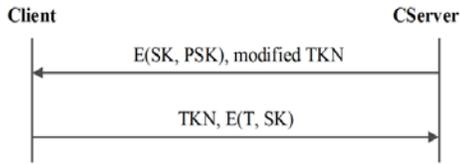

Figure 6. The authentication processes of the proposed protocol

To formally analyze the authentication protocol via SVO, it is important to identify the goal of the analysis. In this protocol, the goal is the following:

**CServer** authenticates the **client**, so that

**CServer** believes (**client** says T)

where PSK is a pre-shared key, and SK is a session key.

Now the protocol analysis will be presented in its five steps, including initial state assumption, received message assumption, interpretation assumption, and derivation.

### 5.1. Initial state assumption

Initial state assumption includes all initial status of the protocol.

I1. *client* believes *client* $\xleftrightarrow{PSK}$ *CServer*

I2. *CServer* believes *client* $\xleftrightarrow{PSK}$ *CServer*

I3. *client* believes (*CServer* controls SK)

I4. *CServer* believes fresh(SK)

I5. *client* believes fresh(SK)

### 5.2. Received message assumption

Received message assumption step indicates what messages each party receives.

R1. *client* received $\{SK\}_{PSK}$

R2. *CServer* received $\{T\}_{SK}$

### 5.3. Comprehension assumption

This step states what receivers believes and indicates what parts of the received message are unknown.

C1. *client* believes (*client* received $\{SK\}_{PSK}$)

C2. *CServer* believes (*CServer* received $\{T\}_{<SK>*CServer}$)

### 5.4. Interpretation assumption

Interpretation assumption step shows what the sender meant by sending the message.

P1. *client* believes (*client* received $\{SK\}_{PSK}$ → *client* received $\{SK \wedge fresh(SK)\}_{PSK}$)

### 5.5. Derivation

Derivation step derives the analysis goal by the previous assumptions.

D1. *client* believes *client* received $\{SK \wedge fresh(SK)\}_{PSK}$

By applying Modus Ponens, C1, P1

D2. *client* believes (*CServer* said SK ∧ *CServer* has SK)

By applying Source Association (A5), D1, I1, I2, and Belief Axiom

D3. *client* $\xleftrightarrow{SK}$ *CServer*

By applying Receiving (A10), R1, D2, and Belief Axiom

D3 shows that *client* and *CServer* communicate with each other by a good shared key (SK).

D4. *CServer* believes (*Client* said T ∧ *Client* has T)

By applying Source Association (A5), C2, D3, and Belief Axiom

D5. *CServer* believes (*Client* said T)

By applying Saying (A16), D4, and Belief Axiom

D6. *CServer* believes (*Client* says T)

By applying Jurisdiction (A21), I5, D5 and Belief Axiom.

D6 shows that our analysis goal, which aimed to prove *CServer* authenticates the *client*, has been achieved by applying the rules of SVO logic.

## 6. Conclusion

DoS attack is currently a major threat and detriment to the availability of cloud services. With each developed defense mechanism against DoS attack, an improved attack appears. Using defense mechanisms with existing authentication protocols to prevent DoS attacks in cloud-computing systems is not always effective. Therefore, this work proposed an authentication protocol to prevent external DoS attacks on cloud-computing systems. Moreover, the security effectiveness of the proposed authentication protocol via SVO logic was analyzed. This analysis proved that the proposed authentication protocol achieves the authentication requirements of an authentication protocol in cloud-computing systems.

# 8. Acknowledgment

This work was partially supported by King Abdulaziz University through the Cultural Bureau of Saudi Arabia in Canada. This support is greatly appreciated.